\documentclass[preprint,5p,sort&compress]{elsarticle}
\usepackage{amsfonts}
\usepackage{amsmath}
\usepackage{amssymb}
\usepackage{graphicx}
\usepackage{sectsty}
\usepackage{bbold}
\usepackage{appendix}
\newcommand\diag{\mathop\mathrm{diag}\nolimits}

\newcommand\be{\begin{equation}}
\newcommand\ee{\end{equation}}
\newcommand\bea{\begin{eqnarray}}
\newcommand\eea{\end{eqnarray}}
\newcommand\bi{\begin{itemize}}
\newcommand\ei{\end{itemize}}
\newcommand\nn{\nonumber}
\newcommand\Ham{\hat{H}}

\newcommand\nr{{\it n}({\bf r})}
\newcommand\kv{{\bf k}}
\newcommand\rv{{\bf r}}

\newcommand{\seq}{$\{P_i\}$}

\newcommand\eqalign[1]{%
	\vcenter{%
		\normalbaselines \advance\baselineskip 5pt
		\advance\lineskip 5pt \tabskip=0pt
		\halign{%
			&\hfil $\displaystyle{##{}}$&
			$\displaystyle{{}##}$\hfil\cr
			#1\crcr
			}%
		}%
	}

\newcommand\dotline{\par\hbox to \hsize{\dotfill}\par}

\makeatletter
\def\befored@t#1.#2.#3;{#1}
\def\afterd@t#1.#2.#3;{#2}
\def\refhead#1{\edef\next{\ref{#1}}\expandafter\befored@t\next..;}
\def\reftail#1{\edef\next{\ref{#1}}\expandafter\afterd@t\next..;}
\def\lsim{\mathrel{\mathpalette\@versim<}}
\def\gsim{\mathrel{\mathpalette\@versim>}}
\def\@versim#1#2{\vcenter{\offinterlineskip
        \ialign{$\m@th#1\hfil##\hfil$\crcr#2\crcr\sim\crcr } }}
\newcommand\becomes[1]{\mathchoice{\becomes@\scriptstyle{#1}}
   {\becomes@\scriptstyle{#1}} {\becomes@\scriptscriptstyle{#1}}
   {\becomes@\scriptscriptstyle{#1}}}
\def\becomes@#1#2{\mathrel{\setbox0=\hbox{$\m@th #1{\,#2\,}$}%
        \mathop{\hbox to \wd0 {\rightarrowfill}}\limits_{#2}}}
\makeatother
\begin{document}
%
%%%%%%%%%%%%%%%%%%%%%%%%%%%%%%%%%%%%%%%%%%%%%%%%%%%%%%%%%%%%%%%%%%%
% Title page
%%%%%%%%%%%%%%%%%%%%%%%%%%%%%%%%%%%%%%%%%%%%%%%%%%%%%%%%%%%%%%%%%%%

\title{Correlations in sequences of generalized eigenproblems arising in Density Functional Theory\tnoteref{t1,t2}}
\tnotetext[t1]{Article based on research supported by the J\"ulich Aachen Research Alliance (JARA-HPC) consortium, the Deutsche Forschungsgemeinschaft (DFG), and the Volkswagen Foundation}
\tnotetext[t2]{Preprints: AICES--2011/08--1 \quad ArXiv:1108.2594}

\author[ads1,ads2]{Edoardo Di Napoli\corref{cor1}}
\ead{e.di.napoli@fz-juelich.de}

\author[ads3]{Stefan Bl\"ugel\corref{cor2}}
\ead{s.bluegel@fz-juelich.de}

\author[ads1]{Paolo Bientinesi\corref{cor2}}
\ead{pauldj@aices.rwth-aachen.de}

\cortext[cor1]{Principal corresponding author}
\cortext[cor2]{Corresponding author}

\address[ads1]{RWTH Aachen University, AICES, Schinkelstr. 2, 52062 Aachen, Germany}
\address[ads3]{Peter Gr\"unberg Institut and Institute for Advanced Simulation, Forschungszentrum J\"ulich and JARA,  52425 J\"ulich, Germany}
\address[ads2]{J\"ulich Supercomputing Centre, Institute for Advanced Simulation, Forschungszentrum J\"ulich, 52425 J\"ulich, Germany}

\begin{abstract}
Density Functional Theory (DFT) is one of the most used ab initio theoretical frameworks in materials science. It derives the  ground state properties of a multi-atomic ensemble directly from the computation of its one-particle density \nr .
In DFT-based simulations the solution is calculated through a chain of successive self-consistent cycles; in each cycle a series of coupled equations (Kohn-Sham)
translates to a large number of generalized eigenvalue problems whose eigenpairs are the principal means for expressing \nr. A simulation ends when \nr\ has 
converged to the solution within the required numerical accuracy. This usually happens after several cycles, resulting in a process calling for the solution of many 
sequences of eigenproblems. 
In this paper, the authors report evidence showing unexpected correlations between adjacent eigenproblems within each sequence. By investigating the numerical properties of the sequences of generalized eigenproblems it is shown that the eigenvectors undergo an ``evolution'' process. At the same time it is shown that the Hamiltonian matrices exhibit a similar evolution and manifest a specific pattern in the information they carry. 
Correlation between eigenproblems within a sequence is of capital importance: information extracted from the simulation at one step of the sequence could be used to compute the solution at the next step. Although they are not explored in this work, the implications could be manifold: from increasing the performance of material simulations, to the development of an improved iterative solver, to modifying the mathematical foundations of the DFT computational paradigm in use, thus opening the way to the investigation of new materials.
\end{abstract}

\maketitle

%%%%%%%%%%%%%%%%%%%%%%%%%%%%%%%%%%%%%%%%%%%%%%%%%%%%%%%%%%%%%%%%%%%%%%%%%
% Content starts here
%%%%%%%%%%%%%%%%%%%%%%%%%%%%%%%%%%%%%%%%%%%%%%%%%%%%%%%%%%%%%%%%%%%%%%%%%

\section{Introduction}
Density Functional Theory~\cite{DG,KS} is a very effective
theoretical framework for studying complex quantum mechanical problems in solid and liquid systems. DFT-based
methods are growing as the standard tools for simulating new materials.
Simulations aim at recovering and predicting physical properties (electronic structure, total energy differences, magnetic properties, etc.) of large molecules as well as systems made of many hundreds of atoms. DFT reaches this result by solving self-consistently a rather complex set of quantum mechanical equations leading to the computation of the {\em one-particle density} \nr %(\ref{eq:dens})
, from which physical properties are derived.

In order to preserve self-consistency, numerical implementations of DFT methods consist of a series of iterative cycles; at the end of each cycle 
a new density is computed and compared to the one calculated in the previous cycle. The end result is a series of successive densities converging to an \nr\ approximating the exact density within the desired level of accuracy~\cite{HK}. Each cycle consists of a complex series of operations: calculation of Kohn-Sham potential, basis functions generation, numerical integration, generalized eigenproblems solution, and ground state energy computation. 
%depending on the basis wave functions that is chosen to represent the orbitals of the quantum systems under examination.
In one particular DFT implementation, namely the Full-potential Linearized Augmented Plane Wave (FLAPW) method~\cite{FKWW,FJ}, matrix entry initialization and generalized eigenvalue problem solution are the most time consuming stages in each iterative cycle (Fig.~\ref{fig:Cyc}). 

The cost for completing these stages is directly related to the number of generalized eigenproblems involved and to their size.
Above a certain threshold, the eigenproblem size is proportional to the third power of the number of atoms of the physical system, while the number of eigenproblems ranges from a few to several hundred per cycle. Typically, each of the problems is dense and a significant fraction of the spectrum is required in order to compute \nr.
The nature of these eigenproblems forces the use of direct methods, such as those included in LAPACK or ScaLAPACK~\cite{LAPACK3,ScaLAPACK}.
All of the most common simulation codes implementing the FLAPW method (WIEN2k, FLEUR, FLAIR, Exciting, ELK~\cite{Wien2k,FLEUR,FLAIR,Exciting,Elk}), despite successfully simulating complex materials~\cite{NIFZC,KFNBB,ADS,CHC}, treat each eigenproblem of the series of iterative cycles in isolation. This implies that no information embedded in the solution of eigenproblems in one cycle is used to speed up the computation of problems in the next. While these routines provide users with accurate algorithms to be used as black-boxes, they do not offer a mechanism for exploiting extra information relative to the application.

The line of research pursued here takes inspiration from the necessity of exploring a different computational approach in an attempt to develop a high performance 
algorithm specifically studied for the FLAPW method. Contrary to the traditional view, we look at the entire succession of iterative cycles making up a simulation as constituted by a few dozen sequences of generalized eigenproblems. By mathematical construction, each problem in a sequence is expected to be, at most, weakly connected to the previous one. At odds with this observation, we present evidence showing that there is an unexpectedly strong correlation between eigenproblems of adjacent cycles in each sequence. We suggest how this extra information should be used to improve the performance of the current state-of-the-art routines. 

Recently some methods have been developed that go in this direction. Among them we mention the block version of the Krylov-Schur~\cite{Zhou-Saad} (in itself an improved version of the Thick-Restart Lanczos~\cite{Wu-Simon}) and the Chebyshev-Davidson~\cite{Zhou} methods. One of the most successful examples in this sense is the recently implemented Chebyshev-Filtered Subspace Accelleration~\cite{Tiago-Cheli} currently included in the PARSEC package specifically targeting ab initio real-space computations.

In Sec.~\ref{sec:Phyfra} we introduce the reader to the DFT framework and more specifically to the FLAPW method. We explain in more detail the 
series of self-consistent cycles, the computational bottlenecks inside each cycle, and how, from this picture, the importance of sequences of eigenproblems emerges.
In Sec.~\ref{sec:Comp}, we illustrate the investigative tools we employ in studying the eigen-sequences, namely eigenvector evolution and unchanging matrix patterns. We then present our computational results and extensively discuss their interpretation from the numerical and physical point of 
view. Finally in Sec.~\ref{sec:corr}  we draw our conclusions and explain how it would be possible to exploit the experimental results to improve the performance of a DFT simulation.

\section{Physical framework}
\label{sec:Phyfra}
%\noindent{\bf Motivation:}\nextline

DFT methods are based on the simultaneous solution of a set of
Schr\"odinger-like equations [eq.~(\ref{eq:KSeq})]. These equations are determined by a Hamiltonian operator $\Ham$ that, in addition to a kinetic energy operator, 
contains an effective potential $v_0[n]$, which functionally depends only on the one-particle electron density $\nr$. In turn, the wave functions $\phi_i({\bf r})$, which 
solve the Schr\"odinger-like equations for $N$ electrons, compute the one-particle electron density [eq.~(\ref{eq:dens})] used in determining the effective potential. The latter is 
explicitly written in terms of the nuclei atomic Coulomb potential $v_I ({\bf r})$, a Hartree term $w({\bf r, r'})$ describing repulsions between pairs of electrons, and 
the exchange correlation potential $v_{xc}[n]({\bf r})$ summarizing all other collective contributions [eq.~(\ref{eq:pot})]. This set of equations, also known as Kohn-Sham (KS)~\cite{KS}, is solved self-consistently. In other words the equations must be solved subject to the condition that the effective potential $v_0[n]$ and the 
electron density $\nr$ mutually agree.
\begin{align}
		\Ham \phi_i({\bf r})  = & \left( -\frac{\hbar^2}{2m} \nabla^2 + v_0({\bf r}) \right) \phi_i({\bf r}) = \epsilon_i \phi_i({\bf r}) 
		%\quad ; \quad \epsilon_1 \leq \epsilon_2 \leq \dots 
		\label{eq:KSeq} \\[0.1cm]
		\nr  = & \sum_i^N |\phi_i({\bf r})|^2  \label{eq:dens} \\[0.1cm]
		v_0({\bf r})  = & v_I({\bf r}) + \int w({\bf r, r'}) n({\bf r'})d{\bf r'} + v_{xc}[n]({\bf r})  \label{eq:pot}
\end{align}

Computational implementations of DFT depend on the particular modeling of the effective potential and on the orbital basis used to parametrize
the eigenfunctions $\phi_i({\bf r})$. In the context of periodic solids, the vector \kv\ and band $\nu$ indices replace the generic index $i$; the Bloch vector $\kv$ is an element of a three-dimensional Brillouin zone discretized over a finite set of values, called the set of $\kv$-points. In the FLAPW method~\cite{FKWW,FJ},
%relies on an augmented plane wave expansion of $\phi_{\kv,\nu}(\rv)$
the orbital function $\phi_{\kv,\nu}(\rv)$ are expanded in terms of a function basis set $\psi_{\bf G}(\kv,\rv)$ indexed by vectors ${\bf G}$ lying in the lattice reciprocal to the configuration space
\be
	\phi_{\kv,\nu}(\rv) = \sum_{|{\bf G + k}|\leq {\bf K}_{max}} c^{\bf G}_{\kv,\nu} \psi_{\bf G}(\kv,\rv).
	\label{eq:comblin}
\ee
In FLAPW, the configuration (physical) space of the quantum sample is divided into spherical regions -- called Muffin-Tin (MT) 
spheres -- centered around atomic nuclei, and interstitial areas between the MT spheres. Within the volume of the solid's unit cell $\Omega$, the basis set $\psi_{\bf G}(\kv,\rv)$ takes a different expression depending on the region
%
%\begin{equation}
%\psi_{\bf G}(\kv,\rv) = \left\{
%	\begin{array}[l]{lr}
%	e^{i({\bf k+G})\rv} & \qquad \textrm{Interstitial}\\
%	\displaystyle\sum_{\it l,m} \left[a^{\alpha,{\bf G}}_{\it lm}(\kv) u^{\alpha}_{\it l}(r) 
%	+ b^{\alpha,{\bf G}}_{\it lm}(\kv) \dot{u}^{\alpha}_{\it l}(r) \right] Y_{\it lm}(\hat{\bf r}_{\alpha}) & \qquad \alpha-\textrm{Muffin Tin}\\
%	\end{array}
%\right.
%\end{equation}
%
%\hspace{-0.5cm}
%
\begin{eqnarray*}
\lefteqn{\psi_{\bf G}(\kv,\rv) = }&  \\
\hspace{-0.5cm} = & \hspace{-0.3cm} \left\{
	\begin{array}[l]{l}
	\frac{1}{\sqrt{\Omega}}e^{i({\bf k+G})\rv} \qquad \qquad \qquad \qquad \qquad \quad -\textrm{Interstitial}\\
	\displaystyle\sum_{\it l,m} \left[a^{\alpha,{\bf G}}_{\it lm}(\kv) u^{\alpha}_{\it l}(r) 
	+ b^{\alpha,{\bf G}}_{\it lm}(\kv) \dot{u}^{\alpha}_{\it l}(r) \right] Y_{\it lm}(\hat{\bf r}_{\alpha}) - \textrm{MT}. \\
	\end{array}
\right.  \\
\end{eqnarray*}

\begin{figure*}[t]
	\hspace{-0.8cm}
	 \includegraphics[scale=0.68]{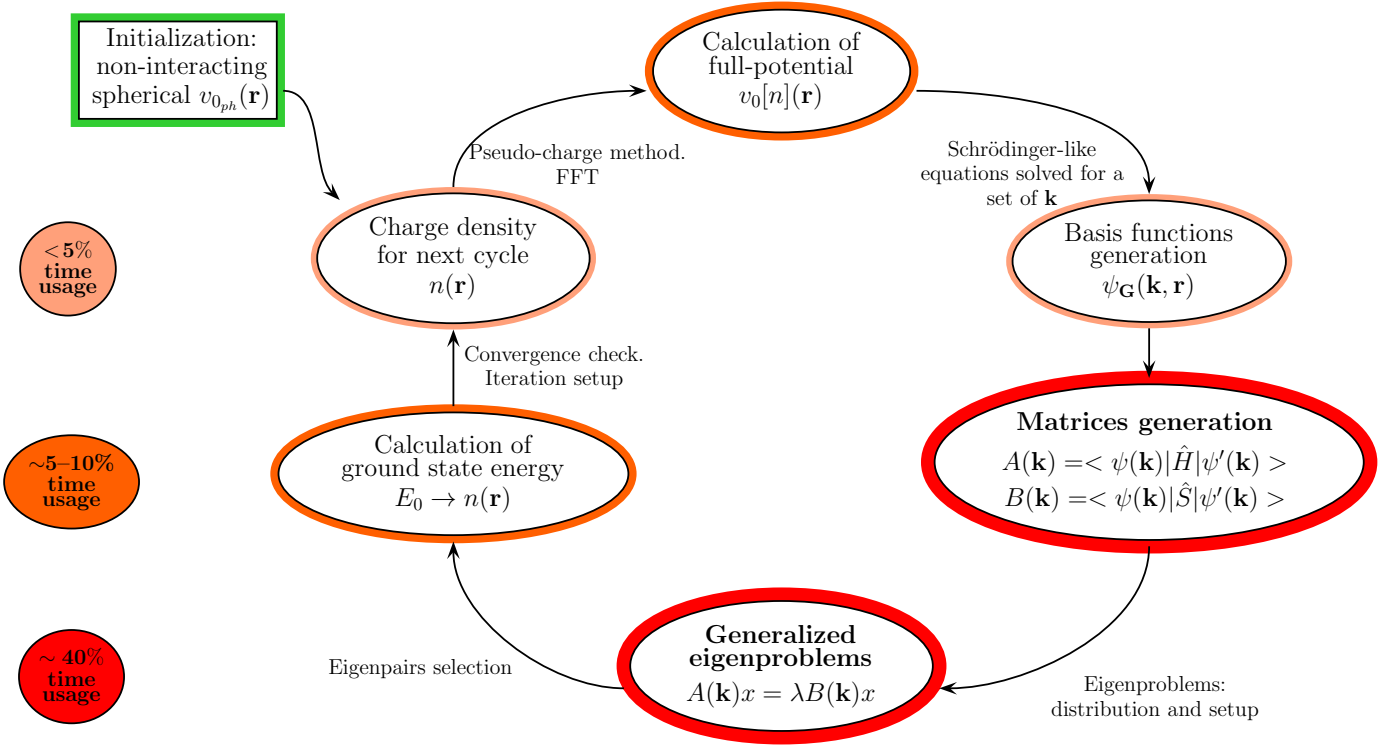}
	\vspace{-0.8cm}
\caption{\small \em A schematic rendering of the principal stages of the self-consistent cycle. Colors indicate the breakdown of the computing time of each stage of the cycle. Each ``time usage'' is based on an average for simulations of very diverse physical systems. Each simulation was run on a single core using a sequential version of the FLEUR code.}\label{fig:Cyc}
\end{figure*}

For each atom $\alpha$, the coefficents $a^{\alpha,{\bf G}}_{\it lm}(\kv)$ and $b^{\alpha,{\bf G}}_{\it lm}(\kv)$ are determined by imposing continuity of the wavefunctions $\phi_{\kv,\nu}(\rv)$ and their derivatives at the boundary of the MT sphere. The $Y_{\it lm}(\hat{\bf r}_{\alpha})$ are spherical harmonics of the $\alpha$-atom, $\hat{\bf r}_{\alpha} \equiv \frac{\rv_{\alpha}}{r_{\alpha}}$ is a unit vector, and $r_{\alpha}$ is the distance from the MT center. The radial functions $u^{\alpha}_{\it l}(r)$ and their time derivatives $\dot{u}^{\alpha}_{\it l}(r)$ are obtained by a simplified Schr\"odinger equation, written for a given energy level $E_l$, containing only the spherical part of the effective potential $v_{0_{ph}}(r)$
\be
	\hspace{-0.2cm} \left\{ - \frac{\hbar^2}{2m} \frac{\partial^2}{\partial r^2} + \frac{\hbar^2}{2m} \frac{l(l+1)}{r^2} + v_{0_{ph}}(r) - E_l \right\}r u^{\alpha}_{\it l}(r) = 0. 
\label{eq:simpSchr}
\ee

Thanks to this expansion, the KS equations naturally translate to a set of generalized eigenvalue problems
\be
\sum_{\bf G'} \left[ A_{\bf GG'}(\kv) - \lambda_{\kv\nu} B_{\bf GG'}(\kv) \right] c^{\bf G'}_{\kv\nu} =0
\label{eq:eigcoef}
\ee
where the coefficients of the expansion $c^{\bf G'}_{\kv\nu}$ are the eigenvectors, while the Hamiltonian and overlap matrices $A$ and $B$ are given by volume integrals and a sum over all MT spheres
\be
	\{A(\kv),B(\kv)\} = \sum \int   \psi^{\ast}_{\bf G}(\kv,\rv) \{\Ham,\hat{\mathbb{1}}\}  \psi_{\bf G'}(\kv,\rv).
	\label{eq:entry} 
\ee

In practical numerical computations, a solution is reached by setting up a multi-stage cycle (Fig.~\ref{fig:Cyc}). An initial educated guess for $\nr$ is used to calculate 
the effective full-potential $v_0[n]$ using the Pseudo-Charge~\cite{W} in combination with Fast Fourier Transform (FFT) methods. The potential, in turn, is inserted into the simplified Schr\"odinger equation (\ref{eq:simpSchr}) whose solutions, together with the coefficents $a^{\alpha,{\bf G}}_{\it lm}(\kv)$ and $b^{\alpha,{\bf G}}_{\it lm}(\kv)$, lead to the basis functions $\psi_{\bf G}(\kv,\rv)$. The latter are used to calculate the entries of the matrices $A(\kv)$ and $B(\kv)$, an operation that requires the 
computation of three-dimensional spherical integrals [eq.~(\ref{eq:entry})] for the non-spherical part of the potential appearing in $\Ham$.
 
In the next stage the matrices just computed are the input in dozens to hundreds of generalized eigenvalue problems~[eq.~(\ref{eq:eigcoef})] that are solved simultaneously. Each eigenproblem is of the form $Ax = \lambda Bx$, where both $A$ and $B$ are dense hermitian matrices, $B$ is additionally positive definite, and x and $\lambda$ form a sought-after eigenpair.
%where $A$ is a dense hermitian matrix, $B$ is dense and positive definite, and x and $\lambda$ form a sought-after eigenpair. 
In FLAPW-related applications, usually only a fraction of the lower part of the spectrum is computed and retained based on the Fermi energy value. The stored eigenpairs are then used to evaluate the ground state energy of the physical system, a step which is followed by the computation of a new charge density $n'({\bf r})$. 

%This section of the cycle can be quite resource consuming
%This is the second most expensive part of the cycle accounting for 40\% of the total computational time. 
At the end of the cycle, convergence is checked by comparing $n'({\bf r})$ with $\nr$. If $| n'({\bf r}) - \nr| > \eta$, where $\eta$ is the required accuracy, a suitable mixing of the two densities is selected as a new guess, and the cycle is repeated. This process is properly referred to as an outer-iteration of the DFT self-consistent cycle. Convergence is guaranteed by the Hohenberg-Kohn theorem~\cite{HK} stating that there exists a unique electron density $n_0({\bf r})$ locally minimizing an energy functional $E[n]$ closely related with the Hamiltonian operator $\Ham$.

In sum, the FLAPW self-consistent scheme is formed by a series of outer-iterations, each one containing multiple large generalized eigenproblems. In order to numerically compute the charge density $\nr$ at each iteration, the matrices $A$ and $B$ need to be initialized for each $\kv$-point and the generalized eigenproblem $Ax = \lambda Bx$ solved. These two stages are the most machine-time consuming part of the cycle, each accounting for between 40\% and 48\% of the total computational time (Fig.~\ref{fig:Cyc}). Moreover, the more complex the material, the larger the matrices and the slower the convergence, resulting in 
an increase in the number of outer-iterations.

\section{An alternative viewpoint}
\label{sec:Comp}

The results presented in this paper originate from the deliberate choice of studying the DFT self-consistent cycle from a different perspective. The entire outer-iterative process is regarded as a set of sequences of eigenproblems $P_i$.
% : it starts from an initial charge density $n_0({\bf r})$, proceeds with a series of iterations $P_1 \dots 
%P_i P_{i+1} \dots P_N$ and converges to a final density $n_f({\bf r})$. 
This interpretation is based on the observation that, for each \kv-point, the solution of a problem 
at a certain iteration $P_i(\kv)$ is a prerequisite for setting up the next one $P_{i+1}(\kv)$. 

Considering the single eigenproblems $P_i$ (the \kv\ index is suppressed for the sake of simplicity) to be part of a sequence $\{P_i\}$ can have far-reaching consequences: it might help to unravel correlations among them and ultimately lead to the conception of an entirely different computational approach to solving them as the simulation progresses. Since DFT is one of the most important ab initio electronic structure frameworks, the study of a computational procedure that would lead to high-performance solutions to \seq\ is of crucial importance.

%Sequences \seq\ are studied by analyzing the evolution of the eigenproblems and their solution as each sequence progresses towards convergence. 
In order to study the evolution of the generalized eigenproblems as part of the sequence \seq, we focus our 
attention on the transformation of eigenvectors and the variation of the matrix entries of the Hamiltonian matrix $A$ (the same could be done for the overlap matrix B). 
For a fixed \kv, each eigenvector at iteration $i$ is compared with its corresponding eigenvector at iteration $i+1$. A similar comparison is performed between the 
values of the entries of adjacent 
%(in the sense of increasing iteration index) 
Hamiltonian matrices. Despite the apparent simplicity of the strategy, the realization of a comparative tool that quantitatively describes the evolution of \seq\ is not a trivial matter. 

\subsection{Eigenvector evolution}
In this section we focus on a generic sequence and describe a procedure to study the evolution of the eigenvectors solving for  $P_i$. The results obtained are independent of \kv\ since each such point represent a vector in the Brillouin zone for which there is an independent sequence of eigenproblems. As a consequence our analysis can be applied to any sequence in the simulation.  
In order to carry out our plan, we need an associative criterion that allows comparison between eigenvectors of successive iterations. This is not a simple task since the ordering of a set of eigenpairs can change substantially from one iteration to the next. 

For instance, one could arrange the eigenvectors by the increasing magnitude of their respective eingenvalues and compare two eigenvectors, say $x^{(i)}_\ell$ and $x^{(i+1)}_\ell$, with the same eigenvalue index $\ell$. This naive comparison is bound to fail due to the fact that eigenvalues close in magnitude often swap positions across iterations. Consequently, identifying eigenvectors becomes rather difficult as the sequence advances. The nature of the self-consistent process interferes with the ability to find a one-to-one correspondence between vectors of neighboring iterations.

\subsubsection{Computational scheme}
\label{sec:comp1}
For a correct comparison between adjacent eigenvectors we developed an algorithm that establishes a one-to-one correspondence based on two observations: 1) a DFT simulation is basically a minimization procedure, and as such favors small eigenpair variations in its progress towards convergence, and 2) all eigenpairs contribute more or less ``democratically'' to the progression of the sequence. Although not fully mathematically rigorous, these statements find their a posteriori justification in the formal comparison between orbitals of two successive iteration cycles (see Appendix~\ref{sec:app}), and translate directly into two specific behaviors of the eigensolutions. First, scalar products between an eigenvector $x^{(i)}_j$ at iteration $i$ and any of the eigenvectors $x^{(i+1)}_\ell$ at iteration $i+1$ 
have a gaussian distribution narrowly peaked at around one value $\sigma^{(i)}_j$. Second, the set of largest scalar products, $\{\sigma^{(i)}_j\}$, has a flat and almost constant distribution. In mathematical terms, they can be written as
\begin{align}
	\forall\ j \ \exists !\ \bar\ell : & \quad  \sigma^{(i)}_j \doteq \langle x^{(i)}_j, x^{(i+1)}_{\bar{\ell}} \rangle \gg \left. \langle x^{(i)}_j, x^{(i+1)}_\ell \rangle \right|_{\ell \neq \bar\ell} \\
	\forall (j_1,j_2) : & \quad \frac{\left|\sigma^{(i)}_{j_1} - \sigma^{(i)}_{j_2}\right|}{\sigma^{(i)}_{j_1} + \sigma^{(i)}_{j_2}} \ll 1. \label{eq:democ}
\end{align}

These observations motivated the design of a routine that, without claiming to be unique or optimized, succeed to correctly relate two successive eigenvectors. Specifically, it identifies, for each eigenvector $x^{(i)}_j$, the largest scalar product $\sigma^{(i)}_j$ subject to the condition that the (i+1)-iteration index $\bar\ell$, associated with $j$, is not associated with any other $j' \neq j$. By design, this procedure establishes a one-to-one correspondence
%\footnote{Look in the appendix for a detailed description of the algorithm --- PDJ help} 
between the eigenvectors of successive iterations, $x^{(i)}_j \Leftrightarrow x^{(i+1)}_{\bar\ell}$, whose information is stored in a permutation operator $\Pi$
\begin{equation}
\forall\ j,\ \exists !\ \bar{\ell} : \quad j = \Pi(\bar{\ell}) \quad \textrm{and} \quad \forall\ \ell,\ \exists !\ \bar{j} : \quad \ell = \Pi^{-1}(\bar{j}).
\end{equation}
Using $\Pi$, the column positions in the matrix of scalar products $\langle x^{(i)}_j, x^{(i+1)}_{\Pi(\ell)}\rangle$ can be rearranged so as to easily obtain the largest scalar products $\{\sigma^{(i)}_j\}$ from the main diagonal.
%We then re-arrange the eigenvalue index $\ell$ so that each $\ell$ is in the same position as its corresponding $j$, construct a permutation 
%operator $\{j\} = P\{\ell\}$, and finally sort $x^{(i+1)}_\ell$ so as to easily obtain the largest scalar products from the main diagonal of 
%the general matrix of scalar products $\langle x^{(i)}_j, x^{(i+1)}_{P(\ell)}\rangle$.
%\be
%	\left(\sigma^{(i)}_{\kv,j_1}, \sigma^{(i)}_{\kv,j_2}, \dots, \sigma^{(i)}_{\kv,j_N}\right) = \diag \left(\langle x^{(i)}_{\kv,\lambda_j}, x^{(i+1)}_{\kv,P(\lambda_\ell)}\rangle
%\right)
%\ee
From this matrix we can easily extract the subspace deviation angles, automatically normalized to one, between corresponding eigenvectors of adjacent iterations 
\be
\theta^{(i)}_j= \diag\left(\mathbb{1} - \langle x^{(i)}_j, x^{(i+1)}_{\Pi(\ell)}\rangle\right).
\ee
These angles provide the means for studying the evolution of the eigenvectors of the sequence of generalized eigenproblems \seq . 

Collecting all the angles computed in one simulation results in a large set of data (there are $N=\dim (A)$ angles for each iteration and each \kv). For our statistical 
analysis, we manipulate the angles so as to plot them in three different ways depending on which parameter characterizing the data is kept fixed. First, fixing the 
iteration index and a specific eigenvalue, we look at how the angles are distributed among the \kv s. Then, we choose a random \kv\ and look at how all the deviation angles 
vary as the sequence progresses. Finally we select an eigenvalue and examine the evolution of the angles for all \kv\ as the iteration index increases.

In order to perform the entire computational process, from eigenvector pairing to deviation angle plotting, we built a Matlab analysis toolkit. The input is the set of matrices $A$ and $B$ of all the eigenproblems appearing in the sequences of a simulation. Simulations of the physical systems analyzed were performed using the FLEUR code~\cite{FLEUR} running on JUROPA, a powerful cluster-based computer operating in the Supercomputing Center of the Forschungszentrum J\"ulich. For each physical system studied we produced outputs for a consistent range of parameters.

\begin{table}[ht]
\caption{ Simulation data}
\centering
\begin{tabular}{c c c c }
\hline \hline%
Material & \shortstack{\# of \\ \kv-points} & \shortstack{\# of \\Iterations} & \shortstack{Avg size of \\matrices}\\ [0.5ex]
\hline%
Fe$_{5\ell}$ & 15 & 27 & ~400\\ %\hline%
ZnO & 40 & 9 & ~490\\ %\hline%
CaFe$_2$As$_2$ & 15 & 30 & ~2600\\
\end{tabular}\\
\label{tab:sim}
\end{table}

\subsubsection{Experimental evidence}

We present here a numerical study for two typical physical systems. The first one is a 5 layer film of iron, denoted by Fe$_{5\ell}$, with a (100) surface orientation modeled by a simple tetragonal lattice containing 5 atoms in the unit cell embedded in two semi-infinite vacua. The second example, zinc oxide, is an ionic bonded material arranged on a wurtzite lattice -- a multilayered hexagonal lattice with 2 Zn and 2 O atoms per unit cell. For each material we ran a 
simulation whose specifics are described in Table \ref{tab:sim}.

\begin{figure}[h]
\hspace{-0.8cm}
\begin{tabular}{c}
	 \includegraphics[width=10.2cm]{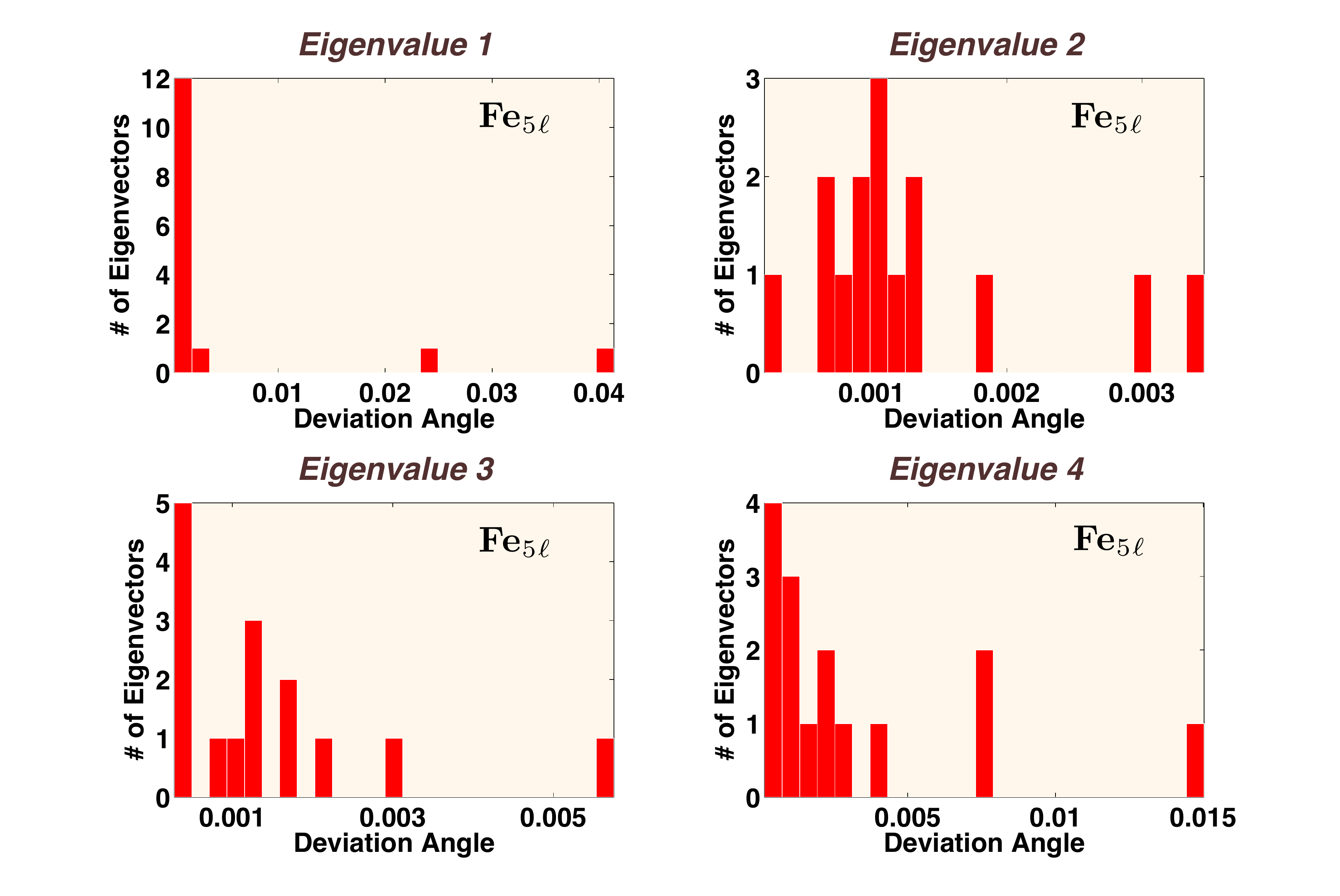}\\
	 \includegraphics[width=10.2cm]{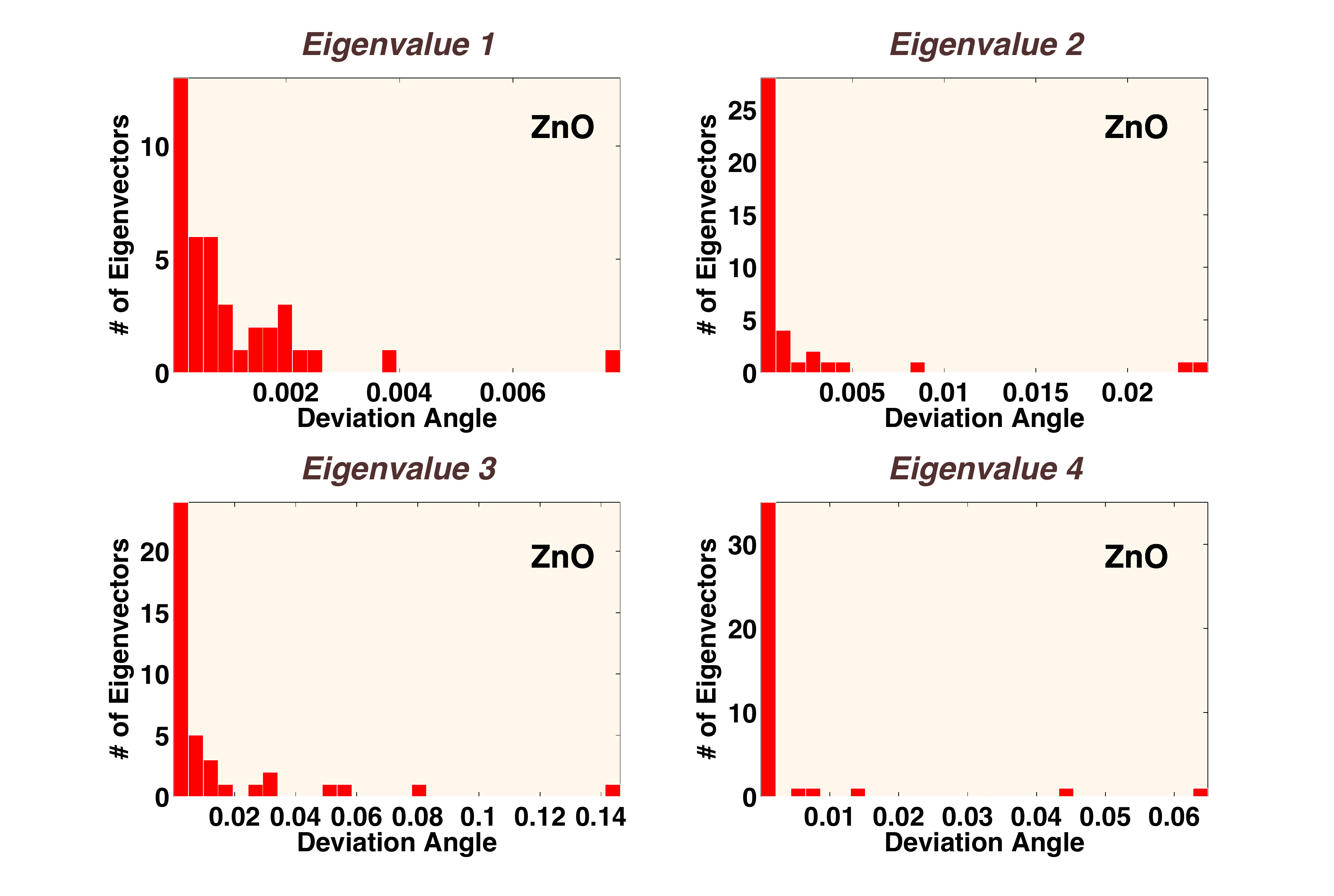}
\end{tabular}
\caption{\small \em Histograms showing a qualitative distribution of the deviation angles of each \kv-point eigenvector corresponding to the four lowest  eigenvalues at a 
fixed iteration. All angles are normalized to one (1.00 = $\pi / 2$). Above: the Fe$_{5\ell}$ case plotted at the $3^{rd}$ iteration. Below: the 
zinc oxide case plotted at the $4^{th}$ iteration.}\label{fig:Hist}
\end{figure}

For each case study we plot a set of histograms showing, for all k-points, the distribution of angle deviations for the four smallest eigenvalues at a specifically chosen iteration (Fig.~\ref{fig:Hist}). In each histogram the distribution is sharply peaked at the lowest end of the interval and has null or only negligible tails. This result supports our analysis as to the ``democratic'' contribution of all angles to the progression of the sequence~[eq.~(\ref{eq:democ})].
 
\begin{figure}[h]
%\hspace{-0.8cm}
%\centering
\begin{tabular}{c}
	 \includegraphics[width=9cm]{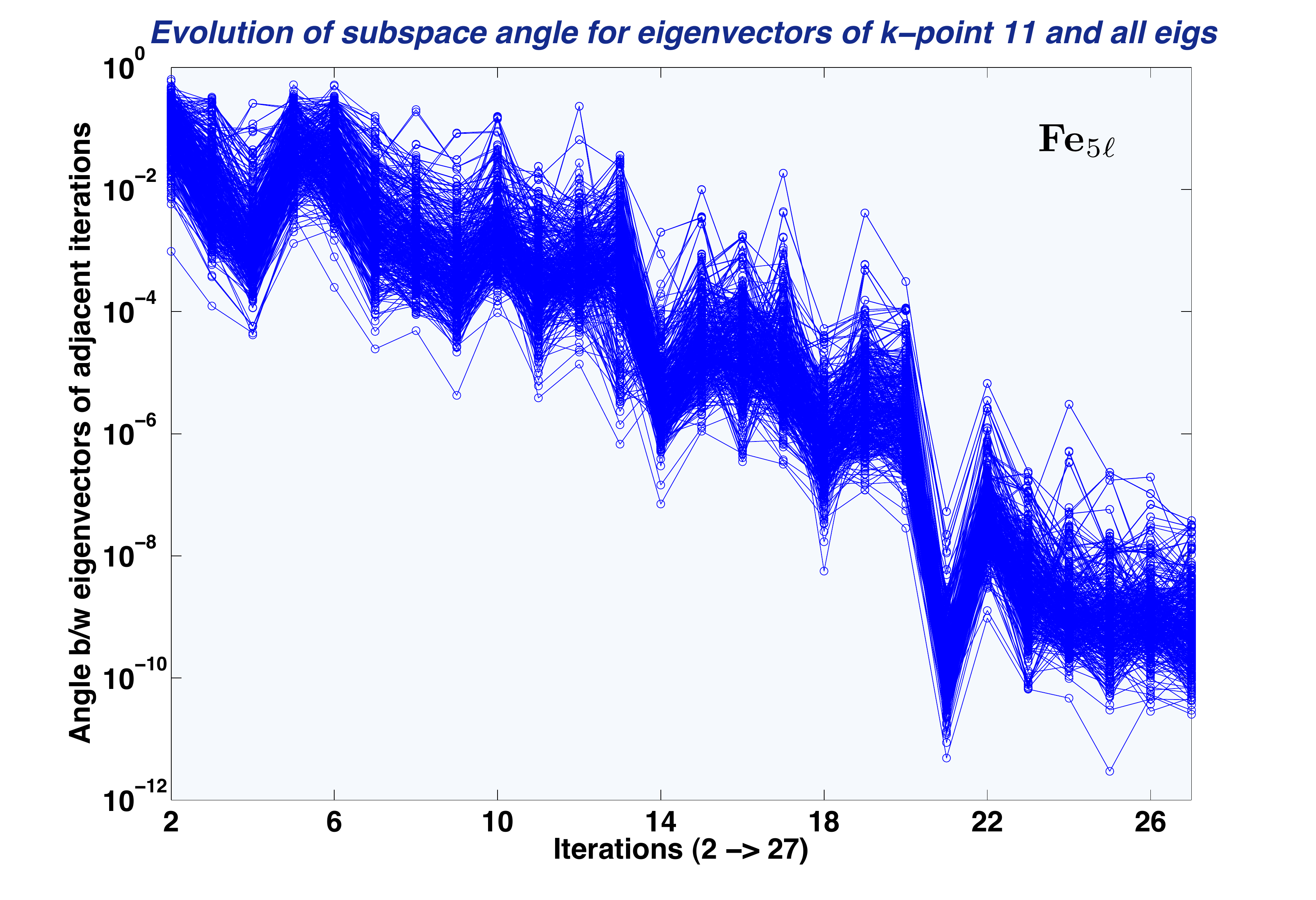} \\  
	 \includegraphics[width=9cm]{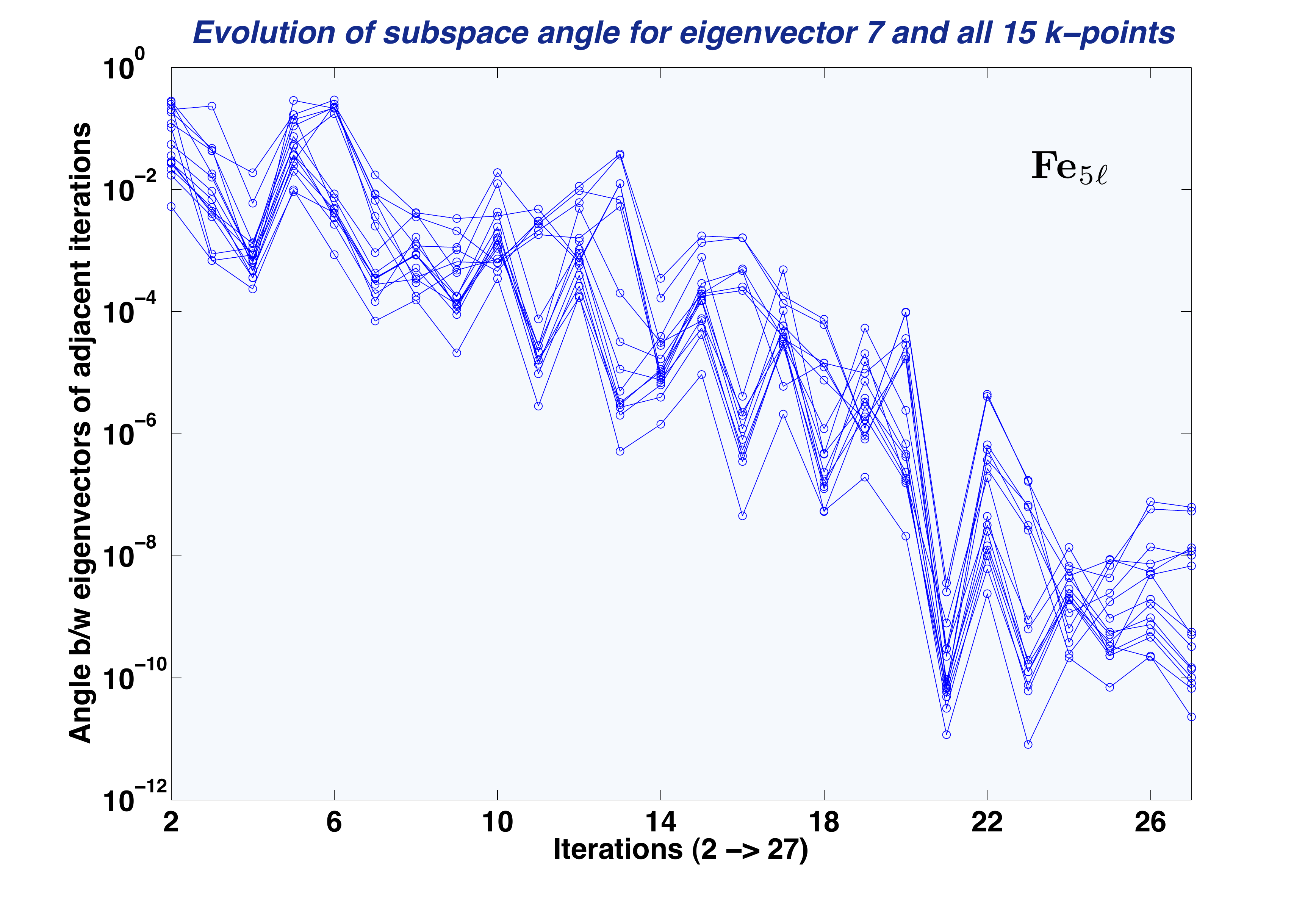} \\  
\end{tabular}
\caption{\small \em Eigenvector angles of successive iterations for the iron multi-layer: 
evolution of angles for all eigenvectors of the sequence corresponding to \kv-point 11 (above) and the evolution of angles for all 15 $\kv$-points of eigenvectors corresponding to the $7^{th}$ lowest eigenvalue (below). All the angles are normalized to one (1.00 = $\pi / 2$).}\label{fig:plot1}
\end{figure}

\begin{figure}[h]
%\hspace{-0.8cm}
%\centering
\begin{tabular}{c}
	 \includegraphics[width=9cm]{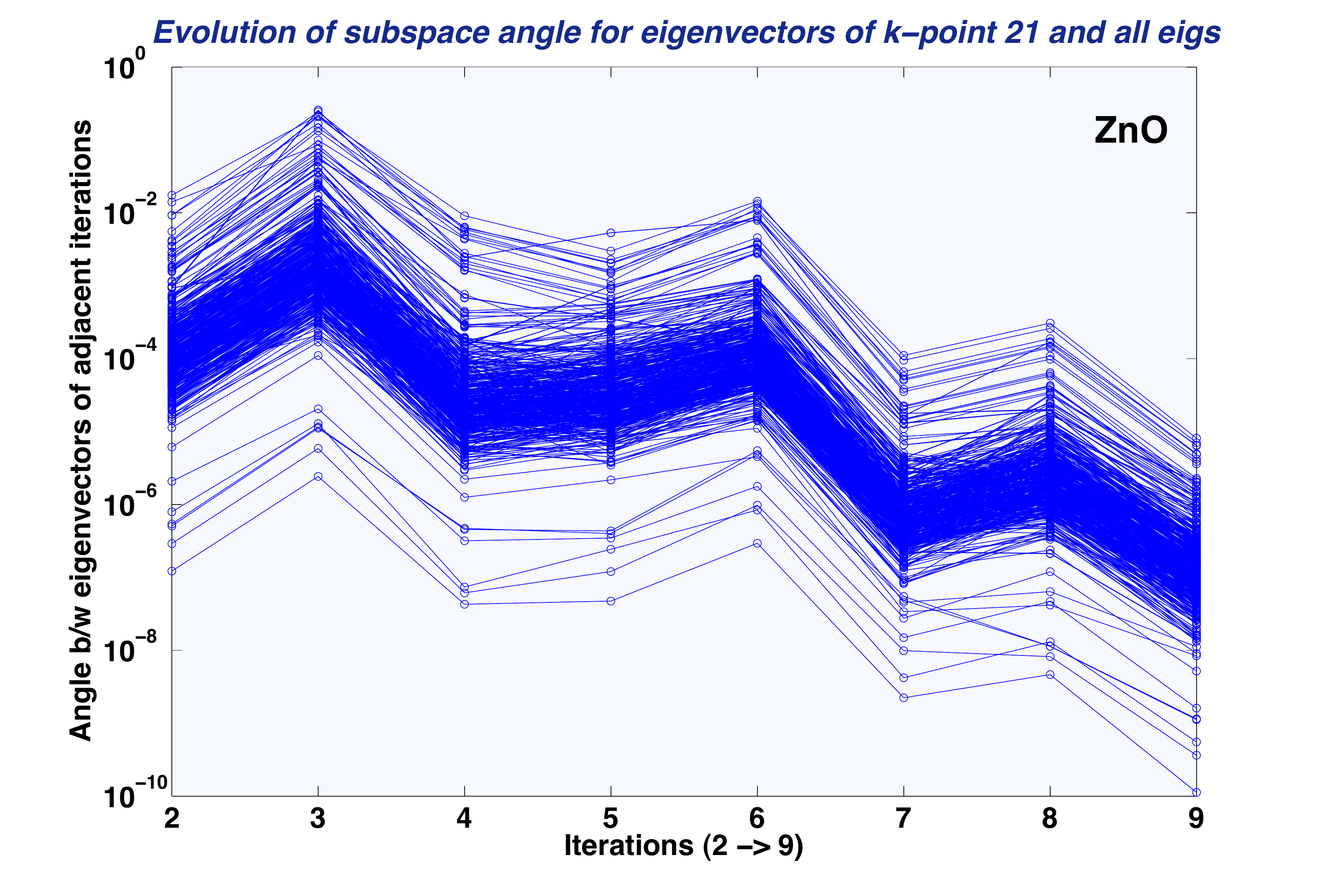}\\
	 \includegraphics[width=9cm]{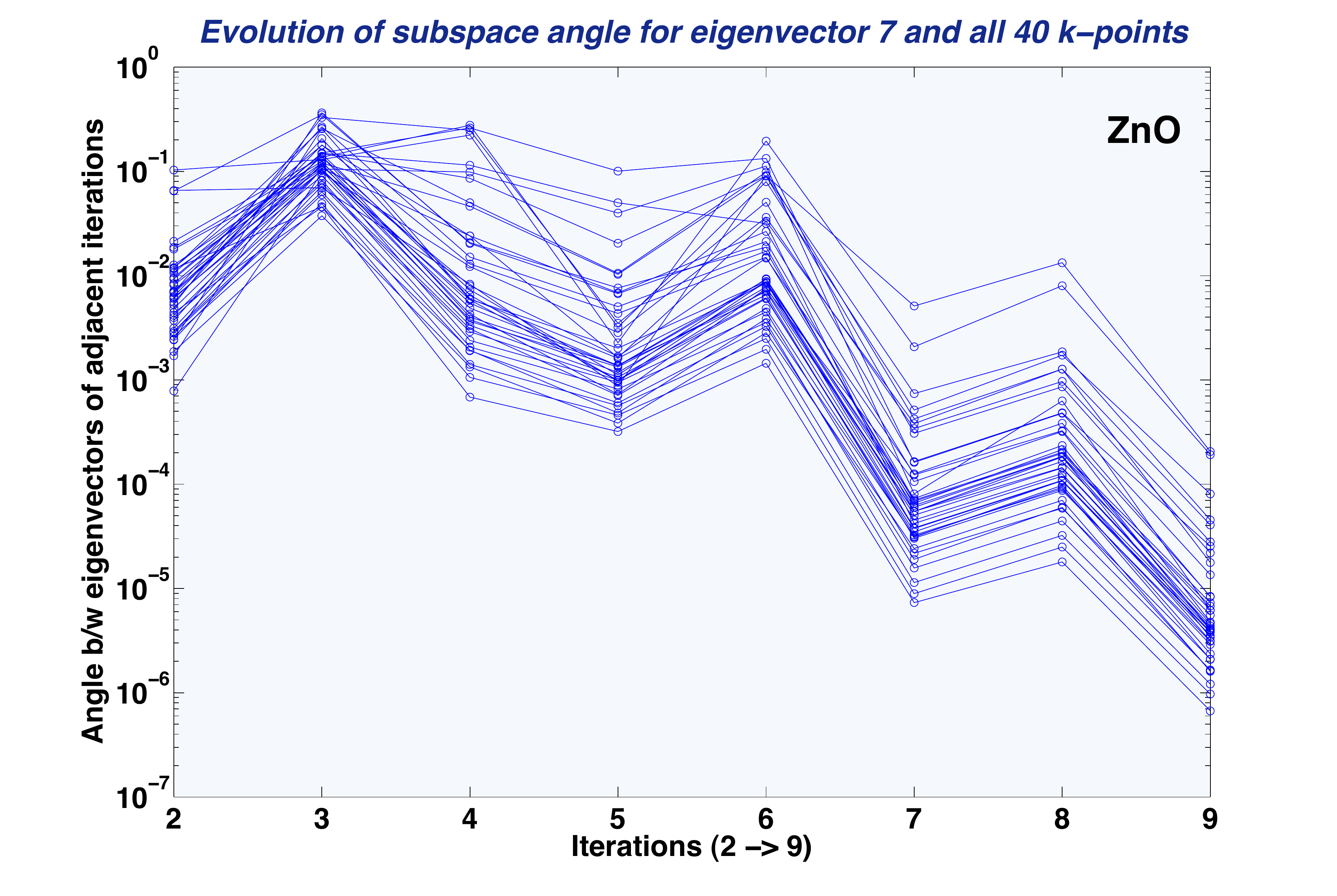}
\end{tabular}
\caption{\small \em Eigenvectors angles of successive iterations for zinc oxide: 
evolution of angles for all eigenvectors of the sequence corresponding to \kv-point 21 (above) and the evolution of angles for all 40 $\kv$-points of eigenvectors corresponding to the $7^{th}$ lowest eigenvalue (below). All the angles are normalized to one (1.00 = $\pi / 2$).}\label{fig:plot1-1}
\end{figure}

For both physical systems we also show two diagrams, one that plots angles for all eigenvectors at one specific \kv-value and the other that presents the angles of the $j^{th}$ eigenvector at each \kv-point (Fig.~\ref{fig:plot1},\ref{fig:plot1-1}). Both graphs are plotted against the iteration index on a semi-log scale to better display the evolution of the deviation angles~\footnote{In all graph labels the abbreviation ``eigs'' stands for the word \textsl{eigenvalues} while the symbol ``b/w'' stands for the word \textsl{between}.} . We can immediately notice the almost monotonic decrease of the deviation angles as the sequences progress towards convergence. Small upward oscillations are probably due to an excess of localized charge that may cause a partial restart of the sequence. We have also observed that the angles corresponding to the lowest 20\% of the spectrum are, on average, higher than the rest. 
%Moreover, as can be seen from the bottom plot of Fig.~\ref{fig:plot1}, some \kv-points, for each selected eigenvalue, have a larger angle evolution suggesting a slightly larger weight in influencing the simulation.

In all other multi-atomic systems studied, besides the ones shown here, the great majority of angles after the $3^{rd}$ or $4^{th}$ iteration are very small. Contrary to intuition the simulation is far from converged at this stage, implying again a sort of ``democracy'' of contribution, where all eigenvectors positively influence the process of minimizing the energy functional that depends on \nr. This behavior has a universal character since we observed it in the bulk, layer, metallic, and ionic materials we analyzed.

In order to give a more quantitative flavor of the eigenvector evolution we have tabulated the mean angle value $\bar{\theta}^{(i)}$ 
%and its standard deviation $\sigma_{\bar{\theta}}^{(i)}$ 
for an iteration at the beginning and one at the end of the simulation (Table~\ref{tab:avg}). Due to their physical relevance, we used only deviation 
angles of those eigenvectors whose eigenvalues represent energies below Fermi level. As can readily be seen, the mean values at the end of the simulation are 
considerably smaller than those at the beginning, in this way confirming the qualitative picture described above.
% and are used to compute the charge density \nr at the end of each iteration.

\begin{table}[ht]
\caption{Deviation Angle Means}
\centering
\begin{tabular}{c c c c c c}
\hline \hline%
Material & \shortstack[c]{{\small \# of relevant eigs} \\ {\small (\% of spectrum)}} & $\bar{\theta}^{\small\rm(start)}$ & $\bar{\theta}^{\small\rm(end)}$ \\ [0.5ex] \hline%
Fe$_{5\ell}$ & 44 (11.0 \%) & 1.10  $10^{-2}$ & 3.64 $10^{-9}$ \\
ZnO & 27 (5.6 \%) & 7.29 $10^{-2}$ & 0.61 $10^{-5}$ \\
\end{tabular}\\
\label{tab:avg}
\end{table}

\subsection{Matrix entry variation}
We systematically look at the variations in the entries in adjacent $A$ matrices for two main reasons. First, we plan to develop a method that identifies those portions of 
entries of $A$ that undergo little or no change at all. Eventually this method could be used to avoid recalculating those entries at each iteration, and in doing so saving computing time. Second, we 
believe that the connection between successive eigenvectors should somehow surface in how much the matrices defining the eigenproblems vary across 
iterations: both are the indirect consequence of changes in the set of basis wavefunctions $\psi_{\bf G}(\kv,\rv)$. %(as explained later in Sec.~\ref{sec:corr}). 

\subsubsection{Computational scheme}
Despite its manifest simplicity, comparison between matrix entries across adjacent iterations can be rather nontrivial. In fact, variations of the single entries of $A_{\kv}^{(i)}$ 
with $A_{\kv}^{(i+1)}$ span a range of several orders of magnitude and need to be opportunely rescaled. Our initial strategy is to normalize all variations so as to map them onto a $[0,1]$ interval. Subsequently we introduce a threshold parameter $p_t$ that cuts off all variations below a certain value. This strategy helps in identifying  those areas of the matrices where the entries undergo relatively large variations; it also allows us to study the percentage of entries that varies as a function of the cut-off value. Eventually, one can determine the value of the threshold that might be chosen for saving computing time while only minimally compromising the accuracy of the eigensolutions (i.e. speed vs accuracy).

First we had to establish the most appropriate metric to gauge the relative size of entry variation. The choice of the metric influences the mapping of the variations onto the specified interval. In this study we chose the maximal entry variation for each matrix difference $\delta^{(i)} = \max(|A_{\kv}^{(i+1)} - A_{\kv}^{(i)}|)$  and normalized each entry of the difference with respect to it. The entries of the resulting matrix $\tilde{A}_{\kv}^{(i)} = \frac{|A_{\kv}^{(i+1)} - A_{\kv}^{(i)}|}{\delta^{(i)}}$ are clearly mapped onto the $[0,1]$ interval. Then the threshold is measured as a fraction of $\delta_{i}$, given by the cut-off value $p_t$, being a number $\in [0,1]$ (e.g. a $p_t=0.1$ value means we are considering all those entries of $\tilde{A}_{\kv}^{(i)}$ that are larger or equal to 10 \% of $\delta^{(i)}$). It has to be noted that, contrary to common intuition, the lower the cut-off value $p_t$, the greater the number of non-zero entries of $\tilde{A}$ is. We complement this analysis with a more conventional approach where the median of the entries of $A_{\kv}^{(i)}$ is compared with the median of the entries of the difference $|A_{\kv}^{(i)} - A_{\kv}^{(i-1)}|$ as the sequence progresses.

All $\tilde{A}_{\kv}^{(i)}$, extracted from the simulation of a given physical system, were analyzed, at a fixed \kv, for different values of the cut-off and for different iteration levels $i$. As for the eigenvector evolution, the input for our analysis is the set of matrices $A$ that defines the eigenproblems appearing in sequences of a 
simulation. All the simulations of the physical systems were performed using the FLEUR code running on JUROPA.

\begin{figure}[h]
\hspace{-0.2cm}
\begin{tabular}{c}
	 \includegraphics[width=8.2cm]{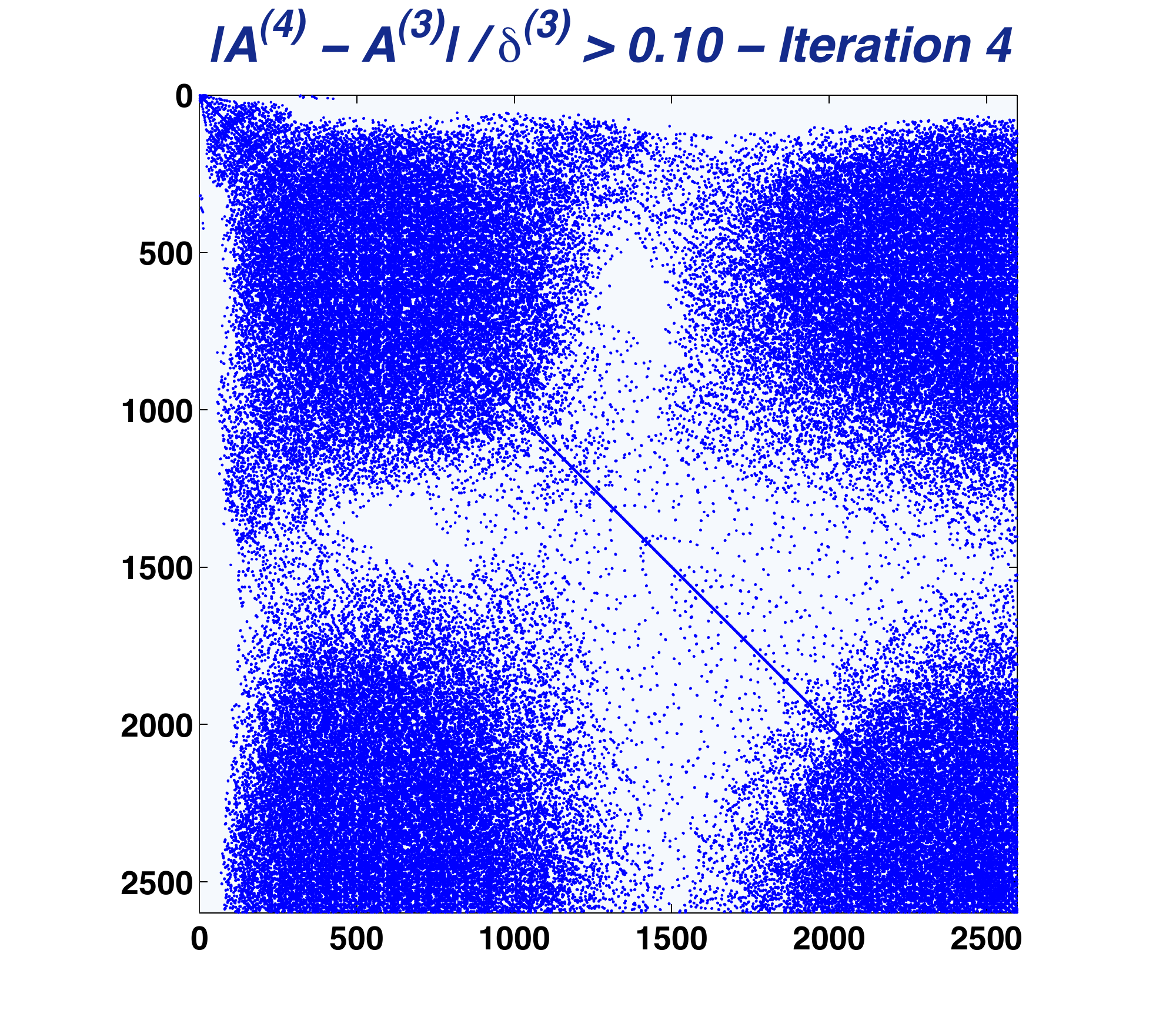}\\
	 \includegraphics[width=8.2cm]{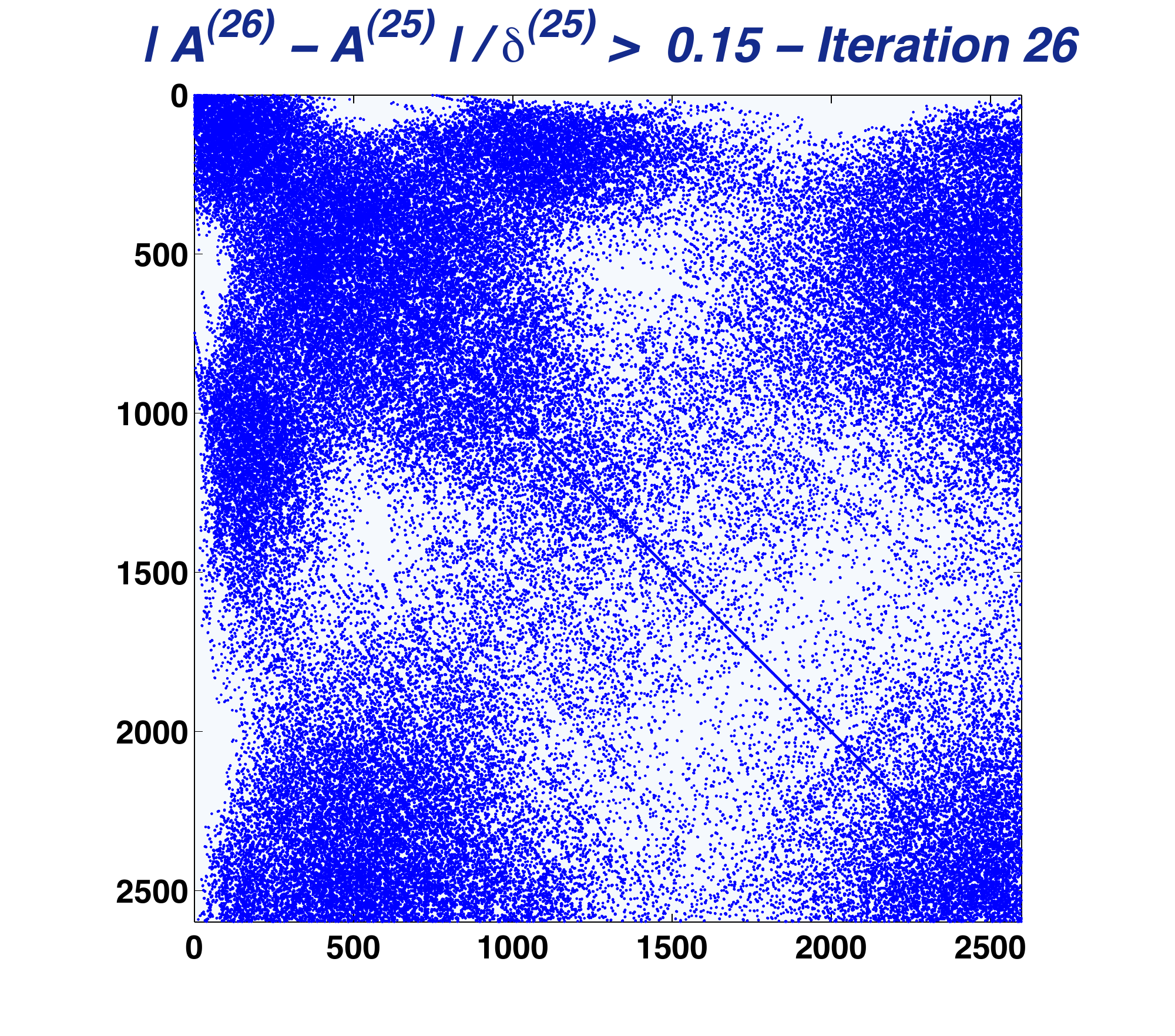}
\end{tabular}
\caption{\small \em Visualization of $\tilde{A}_{\kv}^{(i)}$ for $\kv =1$ at two different iteration cycles, excerpted from a CaFe$_2$As$_2$ simulation. Above: plot of all the entries above the 
cut-off value $p_t=0.10$ at iteration $i=4$. Below: plot of all the entries above the cut-off value $p_t=0.15$ at iteration $i=26$.}\label{fig:spy}
\end{figure}   

\subsubsection{Experimental evidence}
\label{sec:expevi}

As an example of our approach we present the analysis of a simulation of a superconducting compound, denoted by CaFe$_2$As$_2$, that undergoes a first order phase transition from a high temperature, tetragonal phase to a low temperature orthorhombic phase. The specific characteristics of this simulation are listed in Table \ref{tab:sim}. 

In Fig.~\ref{fig:spy} we first give a qualitative picture of the portion of $A$ that changes, for a specific \kv-point at two distant iterations and different cut-off values. Two distinct observations can be made: on the one hand, the empty portions of $\tilde{A}$ tend to preserve their shape and position as the cut-off increases. In other words those parts of $A$ that do not vary much seem to follow a specific pattern independent from $p_t$. On the other hand, the percentage of entries that undergo variation does not seem to be affected by the progress of the sequence of iterations.

\begin{figure}[h]
\hspace{-0.2cm}
\begin{tabular}{c}
	 \includegraphics[width=8.2cm]{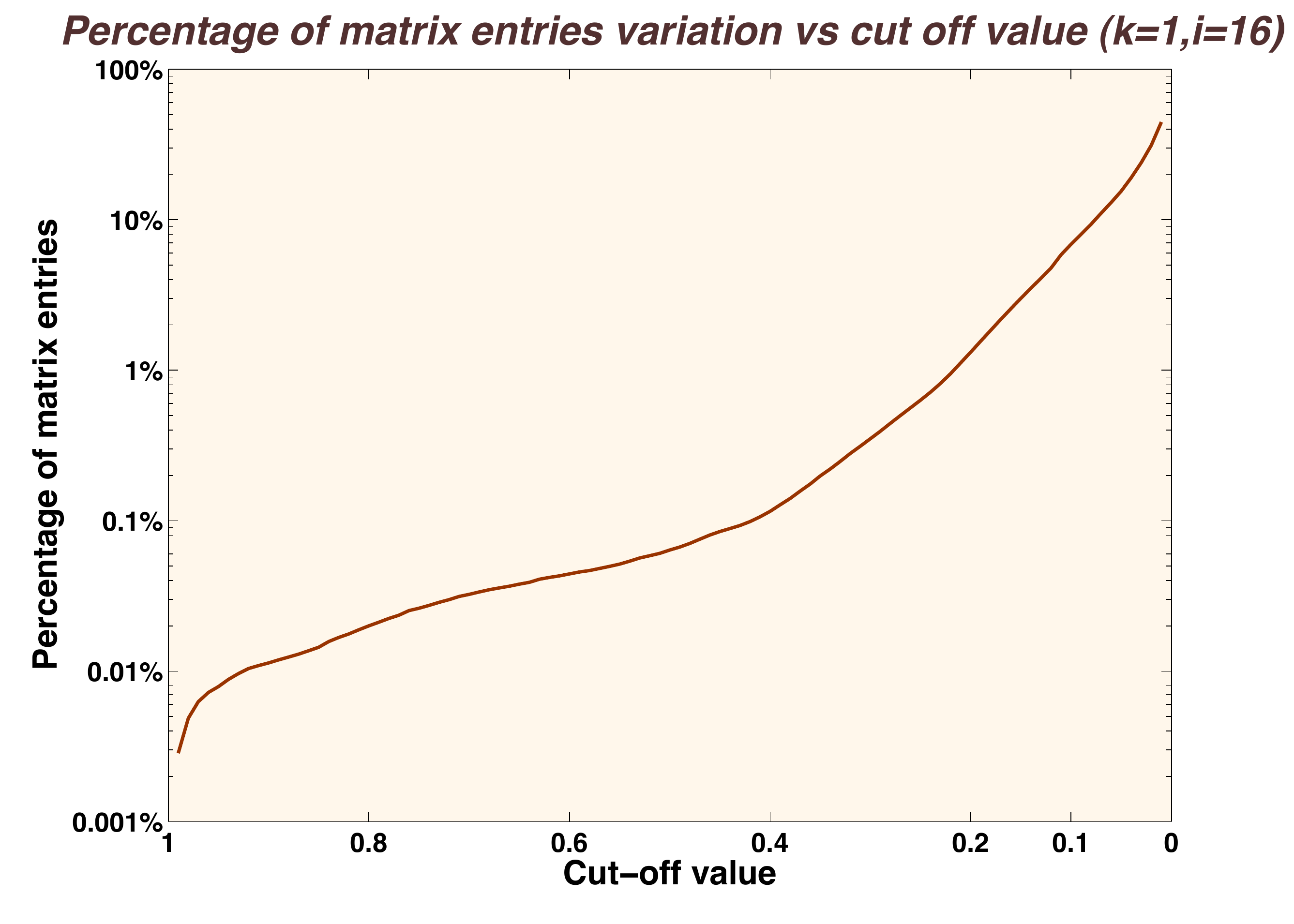}
\end{tabular}
\caption{\small \em Percentage of varying matrix entries plotted versus cut-off values on a semilogarithmic scale for the CaFe$_2$As$_2$ system for $\kv =1$ and $i=16$.}\label{fig:cutoff}
\end{figure}

The fact that such qualitative behavior is evident in all of the physical systems investigated suggests that it is a ``universal'' trait characteristic of DFT-based simulations. This conclusion implies that despite the fact that the basis functions set $\psi_{\bf G}(\kv,\rv)$ changes substantially between 
successive iterations, it would seem that for certain subsets of basis functions such a change contributes very little to the volume integrals in eq.(\ref{eq:entry}).

\begin{figure}[h]
\hspace{-0.2cm}
\begin{tabular}{c}
	 \includegraphics[width=8.2cm]{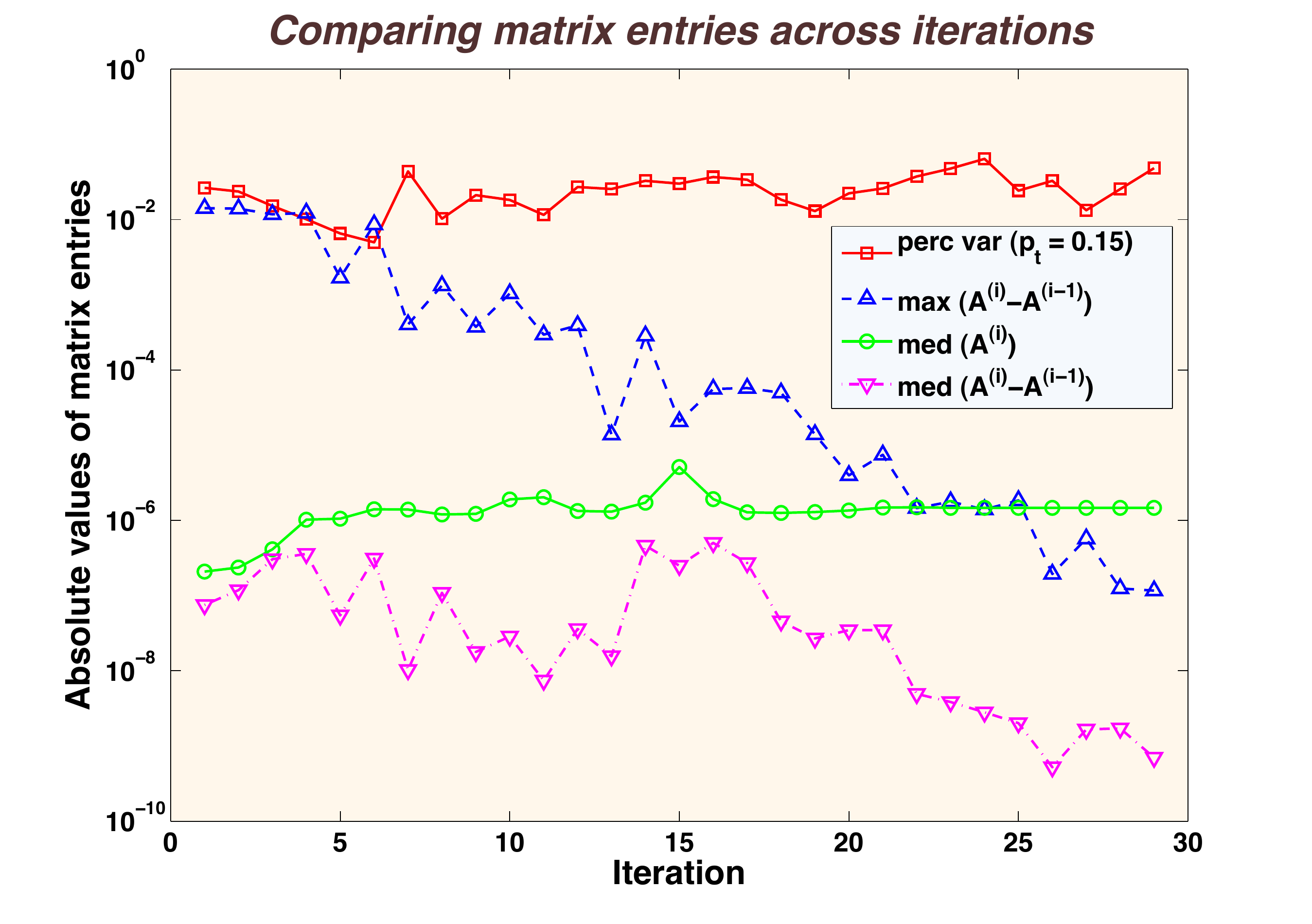}
\end{tabular}
\caption{\small \em Median of the matrix entries of $A_{\bf 1}^{(i)}$ and median and maximum of matrix entries of $|A_{\bf 1}^{(i)} - A_{\bf 1}^{(i-1)}|$ plotted versus the iteration index for the CaFe$_2$As$_2$ system for $\kv =1$.}\label{fig:median}
\end{figure}

In Fig.~\ref{fig:cutoff} we give a more quantitative description, at a fixed iteration, of the number of matrix entries that change as the threshold value increases. It can be noted that the percentage of varied entries becomes significant only for $p_t \leq 0.1$. This behavior indicates that overall there are very few entries undergoing major changes; most of the variations are concentrated in the low end of the metric (i.e. $p_t \times \delta^{(i)}$ and $p_t<0.1$). Contrary to what is suggested by the eigenvector evolution, the percentage of matrix entry variation (at a fixed cut-off) does not seem to decrease as the sequence of eigenproblems progresses (see red line in Fig.~\ref{fig:median}). This is quite a surprising result signaling that unchanging patterns and small eigenvector deviation angles in \seq\ may have different origins.

In Fig.~\ref{fig:median} we have also compared the median and maximum value of $|A_{\kv}^{(i)} - A_{\kv}^{(i-1)}|$ with the median of the entries of $A_{\kv}^{(i)}$. As the simulation progresses the median value of the Hamiltonian matrix remains approximately constant (green line) while the maximum and the median value of the matrix variation decrease. Moreover the median value of $|A_{\kv}^{(i)} - A_{\kv}^{(i-1)}|$ is, on average, 1-2 orders of magnitude lower than $A_{\kv}^{(i)}$ as can also be seen from Table~\ref{tab:median}. 

Can the large number of unchanging entries be used to speed up computations at every DFT iteration? In order to answer this question it needs to be 
understood how the trade off between speed and accuracy depends on the choice of cut-off value in relation to the iteration index $i$. On one hand the patterns and distribution of entries that undergo very small variation could be exploited so as to avoid computing them anew in each iteration. On the other hand the evolution of entry variation suggests that updating some of the entries could be completely avoided after a certain iteration.  

Our study makes manifest that there is room for developing new methods for saving computational time in the process of updating the matrices defining the generalized eigenproblems in \seq . The possible methods and their optimization is still the object of current research and we refer the reader to future publications for more conclusive results. 

\begin{table}[ht]
\caption{Matrix Entry Medians}
\centering
\begin{tabular}{c c c c c c}
\hline \hline%
\shortstack[c]{{\small Iteration} \\ {\small index $(i)$}} & \shortstack[c]{{\small Median} \\ {\small of $A_{\bf 1}^{(i)}$}} &  \shortstack[c]{{\small Median} \\ {\small of $|A_{\bf 1}^{(i)} - A_{\bf 1}^{(i-1)}|$}} & \shortstack[c]{{\small Maximum} \\ {\small of $|A_{\bf 1}^{(i)} - A_{\bf 1}^{(i-1)}|$}} \\ [0.5ex] \hline%
5 & 1.02 $10^{-6}$ & 3.85 $10^{-8}$ & 1.22 $10^{-2}$ \\
10 & 1.22 $10^{-6}$ & 1.76 $10^{-8}$ & 3.74 $10^{-4}$ \\
15 & 1.72 $10^{-6}$ & 1.19 $10^{-7}$ & 2.85 $10^{-4}$ \\
20 & 1.29 $10^{-6}$ & 1.92 $10^{-8}$ & 1.39 $10^{-5}$ \\
25 & 1.47 $10^{-6}$ & 6.31 $10^{-9}$ & 1.40 $10^{-6}$ \\
\end{tabular}\\
\label{tab:median}
\end{table}

\section{Correlation and exploitation}
\label{sec:corr}
In the previous section, we analyzed DFT-based simulations from a non-conventional perspective. Departing from the traditional picture that considers eigenproblems in each iterative cycle of a simulation as independent, we instead assumed that they form a set of sequences of eigenproblems $\{P_i(\kv)\}$. We then provided experimental evidence pointing out a connection between problems that have consecutive iteration indices. In particular, we uncovered a strong correlation between eigenvectors of successive problems as well as the existence of unchanging patterns in the matrices defining the eigenproblems. We illustrated how this correlation is noticeably linked to the convergence process of the simulation: as the iteration index increases, the eigenvector deviation angles become, on average, smaller. Unfortunately we did not observe an equivalent drop in the variation of the matrix entries of the Hamiltonian but only in its maximum and median values. Nonetheless it is evident that the sequences of eigenproblems \seq\ undergo a significant evolution. 

The importance of this result stems from the fact that this correlation is quite unexpected.
Since each single problem $P(\kv)$ at iteration $i+1$ is determined by the orbital wave functions obtained by the solution of all the problems $\{P(\kv)\}$ at iteration $i$, the connection between $P_i$ and $P_{i+1}$ is non-linear and presumably very weak. Until recently it was believed that such weak non-linearity would have hidden any sign of correlation, a conviction that inhibited any research effort in this direction. The source of non-linearity resides in the two indirect ways each $P_i$ is influenced by the basis functions $\psi_{\bf G}(\kv,\rv)$ computed at each new iteration. First, matrix entries defining the eigenproblems are given by volume integrals involving basis functions [eq.~(\ref{eq:entry})], not orbital functions, and second the eigenvectors are the n-tuple of coefficients expressing orbital wave functions in terms of a linear combination of basis functions [eq.~(\ref{eq:comblin})]. As a consequence of these two considerations, eigenvectors were considered to be very loosely connected. The discovery that the contrary to this fact is  true compels us to give great relevance to the evidence of a strong correlation.

While having found a strong correlation between eigenproblems of a sequence \seq\ is in itself an important result, it is even the more so because it opens the way to the exploration of new computational strategies. In particular, the performance of the entire DFT simulation could be improved by boosting the performance of the sequences of eigenproblems. The idea is to take advantage of repetitive patterns in the eigenpencil (Hamiltonian and Overlap) and in the eigenvector evolution. In practice, the key element would be to reuse, for the solution of  eigenproblems at a certain iteration, numerical quantities computed in a previous one. In order to clarify this concept we briefly describe, in the following, the implication of just eigenvector manipulation. 

Traditionally large dense eigenproblems are solved using direct methods whenever the fraction of the sought after spectrum is already above the few \%-points. In the opposite case scenario, iterative methods are mostly used for sparse eigenproblems when the fraction of eigensolutions required is very small. Due to the large number of matrix-vector operations iterative eigensolvers do not perform well  for dense problems and sometimes, in the presence of tight clusters, even fail to converge. These same considerations do not necessarily apply to sequences of dense correlated eigenproblems \seq\ when the fraction of eigenpairs sought after is somewhat lower than 10\%. In this case the reuse of information can somewhat improve the performance of selected iterative solvers. 

Far from presenting any conclusive evidence, we point out that the evolution of eigenvectors could be exploited to improve the performance of iterative solvers and, eventually, make them competitive with direct solvers when applied to sequences of eigenproblems. This result will depend on two distinct key properties of the iterative method of choice: 1) the ability to be fed previously computed eigenvectors and take advantage of them as a starting guess to efficiently compute the new ones, and 2) the capacity to solve simultaneously for a substantial portion of eigenpairs. The first property would result in filtering away as efficiently as possible the unwanted components of the old eigenvectors. The second characteristic implies a moderate-to-substantial speedup for the matrix-vector multiplications as well as an improved convergence. This is part of a study that is underway and will be presented in a future publication.

\section{Conclusions}
The results described in this paper are an example of how it is feasible to study a mathematical model by reverse induction. 
Starting from simulations, we showed how it is possible to construct a method to analyze the potential improvements of the algorithmic realization of a mathematical model on which the simulations are based. This approach reverses the usual direction that goes from theoretical model to experiment passing through computer-based simulations. In other words it is an example of how to look at DFT-based computations as an inverse problem. As such we would like to refer to our approach as a ``reverse simulation'' method.

In this work we utilize the ``reverse simulation'' approach as a tool to investigate the properties of the non-linear generalized eigenproblem arising from the Kohn-Sham equations. In general such an eigenproblem is linearized through the introduction of a self-consistent cycle that is solved until convergence is reached. The non-linear problem translates then into a series of eigenproblems that are solved in total independence from each other. We show that in reality these eigenproblems are strongly correlated and constitute part of a sequence. We finally suggest that the correlation discovered could be exploited by a class of opportunely designed iterative solvers.

This result is of great impact for the community of computational physicists working in the wide field of material science. It could be the first step towards changing computational paradigm, especially in view of the increasing trend toward the use of massively parallel architectures. By giving ``dense'' DFT methods, like FLAPW, access to the use of iterative solvers one will effectively increase their scalability. Conversely a higher scalability would empower these ab initio methods with the capability of investigating larger physical systems that are currently out of reach.

Besides affecting DFT methods that until now have precluded the use of iterative solvers,
%due to their high computational cost and rather poor performance when utilized to solve dense eigenproblems. 
the methodological viewpoint used here will have by far more important consequences than just improving the computational approach to the simulations: it would allow us to go beyond the conventional FLAPW method and create a more efficient mathematical paradigm.

%%\vspace{-0.1cm}
\section{Acknowledgements}
We thank Dr. Daniel Wortmann, Dr. Gustav Bihlmayer and Gregor Michalicek for discussions and their help in dealing with the FLEUR code. The computations were performed under the auspices of the  J\"ulich Supercomputing Centre at the Forschungszentrum J\"ulich, whose generous  support of CPU time is hereby acknowledged. We would also like to thank the AICES graduate school for hosting some of the authors and contributing to the success of the project. 

Financial support from the following institutions is gratefully acknowledged: the JARA-HPC through the Midterm Seed Funds 2009 grant, the Deutsche Forschungsgemeinschaft (German Research Association) through grant GSC 111, and the Volkswagen Foundation through the fellowship ``Computational Sciences". \\
%\vspace{0.5cm}
\begin{flushleft}
      {\bf \Large Appendix}
    \end{flushleft}
\begin{appendices}
\numberwithin{equation}{section}
\section{A formal justification of the eigenvector computational scheme}
\label{sec:app}
The formal study of eigenvector evolution is based on two statements given at the beginning of section 3.1.1. Of these statements the first one simply states the fact that the orbital functions that make up the charge density have to already be a very good guess for the exact wavefunctions for the self-consistent cycle to converge. This fact translates to the following formal analysis.

Eq.~(\ref{eq:comblin}) written for the orbital functions at a certain iteration $(i)$ is
\be
	\phi^{(i)}_{\kv,\nu}(\rv) = \sum_{|{\bf G + k}|\leq {\bf K}_{max}} c^{(i){\bf G}}_{\kv,\nu} \psi^{(i)}_{\bf G}(\kv,\rv).
	\label{eq:comblini}
\ee
If we take the scalar product of two $\phi$s for different band indices but the same \kv-point and the same iteration we straightforwardly obtain
\begin{align}
	& \int d\rv\ \phi^{*(i)}_{\kv,\nu}(\rv) \phi^{(i)}_{\kv,\mu}(\rv) = \nn \\
	= & \sum_{|{\bf G + k}| , |{\bf G' + k}|\leq {\bf K}_{max}} c^{*(i){\bf G'}}_{\kv,\nu}c^{(i){\bf G}}_{\kv,\mu} \int d\rv\ \psi^{*(i)}_{\bf G'}(\kv,\rv) \psi^{(i)}_{\bf G}(\kv,\rv)\nn \\
	= &  \sum_{|{\bf G + k}| , |{\bf G' + k}|\leq {\bf K}_{max}} c^{*(i){\bf G'}}_{\kv,\nu} S^{(i)}_{\bf G' G}\ c^{(i){\bf G}}_{\kv,\mu}\\
	= &\ \langle x_\nu x_\mu \rangle = \delta_{\nu \mu}\nn.
\end{align}
By repeating the same computation for $\phi$s from two successive iterations, we arrive at
\begin{align}
	& \int d\rv\ \phi^{*(i)}_{\kv,\nu}(\rv) \phi^{(i+1)}_{\kv,\mu}(\rv) = \nn \\
	= & \sum_{|{\bf G + k}| , |{\bf G' + k}|\leq {\bf K}_{max}} c^{*(i){\bf G'}}_{\kv,\nu}c^{(i+1){\bf G}}_{\kv,\mu} \int d\rv\ \psi^{*(i)}_{\bf G'}(\kv,\rv) \psi^{(i+1)}_{\bf G}(\kv,\rv)\nn \\
	= &  \sum_{|{\bf G + k}| , |{\bf G' + k}|\leq {\bf K}_{max}} c^{*(i){\bf G'}}_{\kv,\nu} \tilde{S}_{\bf G' G}\ c^{(i+1){\bf G}}_{\kv,\mu}
\label{eq:mixing}
\end{align}
where $\tilde{S}_{\bf G' G}$ is an asymmetric positive definite matrix and as such can be factorized according to Cholesky as
\be
	 \tilde{S}_{\bf G' G} = \sum_{|{\bf G'' + k}|\leq {\bf K}_{max}} \tilde{L}_{\bf G' G''} \tilde{L}^T_{\bf G'' G} 
\ee
with $\tilde{L}_{\bf G' G''}$ being a lower triangular matrix.

Let us momentarily assume the correctness of the statements in section~\ref{sec:comp1}. The first statement implies that the left-hand side of Eq.~(\ref{eq:mixing}) can be expanded as $\langle x^{(i)}_\nu x^{(i+1)}_\mu \rangle = \delta_{\nu\mu} + \epsilon E_{\nu\mu} + o(\epsilon^2)$ with $\epsilon$ being a generic numerical expansion parameter. Correspondingly, from the second statement we conclude that the matrix $E_{\nu\mu}$ should be of quasi-block-diagonal form with a dominant diagonal which enforces only small rotations or mixing. In the same fashion we can arbitrarily expand the $\tilde{S}_{\bf G' G}$ on the right-hand side
\be
	\tilde{L} \tilde{L}^T = L^{(i)} \left( I + \epsilon D + o(\epsilon^2)\right) L^{(i+1)T}.
\ee
where $L^{(i)}$ and $L^{(i+1)}$ indicate the lower triangular matrices decomposing $S^{(i)}_{\bf G' G}$ and $S^{(i+1)}_{\bf G' G}$ respectively.
Combining these two expansions we arrive at the conclusion that 
\begin{align}
\label{eq:exp}
		\langle x^{(i)}_\nu x^{(i+1)}_\mu \rangle = &\ c^{*(i)}_{\kv,\nu} L^{(i)}\ L^{(i+1)T}\ c^{(i+1)}_{\kv,\mu} \\
		= &\ \delta_{\nu\mu} + \epsilon \left[ E_{\nu\mu} - c^{*(i)}_{\kv,\nu}\ L^{(i)}\ D\ L^{(i+1)T}\ c^{(i+1)}_{\kv,\mu} \right] + o(\epsilon^2).\nn
\end{align}
In other words, depending on the size of the numerical parameter $\epsilon$, one may expect small angle variations between eigenvectors of successive iteration cycles. While this result is formally correct, it is based on an unproven assumption. From the analytical point of view very little can be said on the validity of this expansion since each cycle updates the basis functions in a very non-linear manner. We reverse the usual reasoning and by assuming the correctness of the expansion work our way back. In practice, we verify numerically the consistence of Eq.~\ref{eq:exp} and consequently can infer the validity of its premises. 
%One that, from an exhaustive numerical analysis of the behavior of eigenvector angles, can infer the validity of the initial statements. 
In other words we follow an inverse engineering problem approach as the basis for the computational scheme of subsection~\ref{sec:comp1}.
\end{appendices}
%%%%%%%%%%%%%%%%%%%%%%%%%%%%%%%%%%%%%%%%%%%%%%%%
%% BACKMATTER
%%%%%%%%%%%%%%%%%%%%%%%%%%%%%%%%%%%%%%%%%%%%%%%%
\bibliographystyle{plain}
%\vspace{-0.1cm}

\end{document}